\documentclass[twocolumn,letter]{IEEEtran}
\usepackage[usenames,dvipsnames]{color}
\usepackage{subfigure}
\usepackage{graphicx}
\usepackage{amsmath}
\usepackage{color}
\usepackage{multicol}
\usepackage{amssymb}

\usepackage{amsfonts}%
\usepackage{verbatim}%
\usepackage{enumerate}%
\usepackage{psfrag}
\usepackage{epsfig}
\usepackage{multicol}
\usepackage{setspace}
\usepackage{dsfont}
\usepackage{balance}

\usepackage{algorithm}
\usepackage{algorithmic}
\usepackage{cite}
\usepackage{float}
\usepackage{cite}
 
\begin{document}
\title{Comments on ``Optimal Utilization of a Cognitive Shared Channel with a Rechargeable Primary Source Node"}
\author{ Ahmed El Shafie$^\dagger$, Ahmed Sultan$^\star$\\
\small \begin{tabular}{c}
$^\dagger$Wireless Intelligent Networks Center (WINC), Nile University, Giza, Egypt. \\
$^\star$Department of Electrical Engineering, Alexandria University, Alexandria, Egypt.
\end{tabular}
}
\date{}
\maketitle
\begin{abstract}
In a recent paper \cite{pappas2012optimal}, the authors investigated the maximum stable throughput region of a network composed of a rechargeable primary user and a secondary user plugged to a reliable power supply. The authors studied the cases of an infinite and a finite energy queue at the primary transmitter. However, the results of the finite case are incorrect. We show that under the proposed energy queue model (a decoupled ${\rm M/D/1}$ queueing system with Bernoulli arrivals and the consumption of one energy packet per time slot), the energy queue capacity does not affect the stability region of the network.
\end{abstract}
%\vspace{-0.5 cm}
%\begin{IEEEkeywords}
%%\vspace{-0.3 cm}
%Cognitive radio; protocol design; throughput analysis; stability region; queue stability; multiple access.
%\end{IEEEkeywords}

%\vspace{-0.2cm}
\section{Introduction}
%\vspace{-0.2cm}
In a recent paper \cite{pappas2012optimal}, Pappas {\it et al.} considered a cognitive setting with a single primary transmitter-receiver pair and a single secondary transmitter-receiver pair. The primary transmitter is assumed to be a rechargeable (battery-based) node, whereas the secondary transmitter is assumed to be plugged to a reliable power supply. Each transmitter has a data buffer with infinite capacity for storing its incoming traffic. The primary transmitter has an additional energy queue with buffer capacity of $c$ packets for harvesting the energy packets from the environment \cite{pappas2012optimal}. To render the characterization of the stability region tractable, the authors assumed that the energy queue is modeled as ${\rm M/D/1/c}$ queue and expends one energy packet each time slot regardless of the primary data queue state and the rest of the queues in the system. Hence, the energy queue is totally {\it decoupled} of the other queues in the system. The authors investigated both cases of an infinite ($c\rightarrow\infty$) and a finite ($c<\infty$) energy queue.
 %In System Model and Section V of \cite{pappas2012optimal}, the primary energy queue, queue $B$, is assumed to be modeled as a decoupled ${\rm M/D/1/c}$ with an energy packet consumption each time slot and Bernoulli arrivals, which implies that the energy queue service rate is deterministic with rate $1$ energy packet per time slot (an energy packet consumption each time slot regardless of the other queues) and the maximum number of arrivals to $B$ per time slot is only one energy packet.
 However, equation (32) in \cite{pappas2012optimal} is for a continuous-time ${\rm M/M/1/c}$ queueing system with Poisson arrivals and with service rate equals to $1$ energy packet consumption per time slot. Therefore, the authors used an incorrect formula for the probability of the energy queue being nonempty \cite[Eqn. (32)]{pappas2012optimal}. Consequently, all results and conclusions in \cite[Sec. V]{pappas2012optimal} and the plot of finite energy buffer in Figs. 2 and 3 are incorrect.
%%This is because as mentioned
More specifically, for the finite capacity primary energy queue, in \cite[Sec. V]{pappas2012optimal}, the energy queue was mentioned to be modeled as a decoupled discrete-time ${\rm M/D/1/c}$ system with Bernoulli arrivals of rate $\delta$ and service rate of one packet consumption per time slot. The authors of \cite{pappas2012optimal} mentioned that the probability of the energy queue $B$ being nonempty is
\begin{equation}
{\rm Pr}\{B\ne0\}=\frac{\delta(1-\delta^c)}{1-\delta^{c+1}}
\label{tot}
\end{equation}
However, this formula is for a continuous-time queue modeled as ${\rm M/M/1/c}$ with unity service rate and Poisson arrivals.\footnote{The formula in (\ref{tot}) and the analysis of ${\rm M/M/1/c}$ queueing system can be found in many references such as \cite[page 158]{zoz}, \cite[page 123]{ibe2008markov}, \cite[page 427]{stewart2009probability}, \cite[page 424]{ross2000introduction}.} This contradicts the assumption of discrete-time ${\rm M/D/1/c}$ queue with Bernoulli arrivals. We provide here the correct formula of the energy queue being nonempty, which is straightforward to derive.

  \begin{figure}
  \centering
  \includegraphics[width=1\columnwidth]{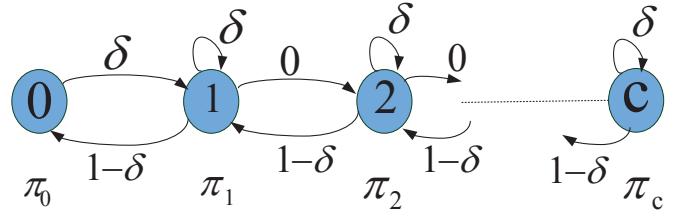}\\
  \caption{Markov chain modeling the primary energy queue $B$. The transition from $j\in\{0,1,2,\dots,c\}$ to $j+k$, where $k\in\{2,3,\dots\}$, is zero because the arrivals are Bernoulli which implies a maximum of one packet arrival per time slot, while one energy packet is consumed per slot if the energy queue in nonempty and even if the data buffer is empty. In the figure, $\pi_j$, $j\in\{0,1,2,\dots,c\}$, represents the probability of the energy queue having $j$ energy packets.
  }\label{fig1}
  \end{figure}

  %\cite{giambene2005queuing}
Since the arrivals to the energy queue are Bernoulli with rate $\delta$, the departure rate is deterministic with rate $1$ energy packet per time slot, and the queue is decoupled of all other queues in the system; the Markov chain of the energy queue can be modeled as in Fig. \ref{fig1}. The transition probability from state $0$ to state $0$ is $1\!-\!\delta$, from state $0$ to state $1$ is $\delta$, from state $1$ to state $0$ is $1\!-\!\delta$, from state $1$ to state $1$ is $\delta$, and the rest of the transition probabilities are zero. We note that the probability of moving from state $j\in\{0,1,2,\dots,c\}$ to state $j+k$, where $k\in\{2,3,\dots\}$, is equal to zero due to the Bernoulli arrival assumption as mentioned earlier.
Solving the state balance equations of the Markov chain, the probability that the energy queue being in state $0$ and state $1$ are  $\pi_0\!=\!1-\delta$ and $\pi_1\!=\!\delta$, respectively. Therefore, regardless of the buffer size, the Markov chain has only two stationary states with non-zero probabilities, namely state 0 where the queue is empty and its steady state probability is $\pi_0\!=\!1\!-\!\delta$, and state $1$ where the queue has one packet and its steady state probability is $\pi_1\!=\!\delta$.

Note that the Markov chain of the primary energy queue with general service rate $0\!\le\!\mu_{\rm e}\!\le\! 1$ can be modeled as in \cite[page 119]{giambene2005queuing}. Setting the notations in \cite[page 119]{giambene2005queuing} with $a=\delta$, $b\!=\!\mu_{\rm e}\!=\!1$ and $K\!=\!c$, we will get exactly the formulas of steady-state probabilities provided in this comment.

Finally, we conclude that the buffer size under such trivial model will not affect the stability region. That is, the cases of infinite and finite buffer capacity of the energy queue are equivalent in terms of nodes' throughput and result in the same stability region.
\balance
\bibliographystyle{IEEEtran}
\bibliography{IEEEabrv,term_bib}
\end{document}